\documentclass[sigconf]{acmart}

\usepackage{booktabs}
\usepackage{graphicx}
\usepackage{tabularx}

\renewcommand\footnotetextcopyrightpermission[1]{}
\begin{document}

\title{Allow Me Into Your Dream: A Handshake-and-Pull Protocol for Sharing Mixed Realities in Spontaneous Encounters}

\author{Botao Amber Hu}\authornote{These authors contributed equally to this work.}\authornote{Corresponding author}
\orcid{0000-0002-4504-0941}
\affiliation{
  \institution{Reality Design Lab}
  \city{New York City}
  \country{USA}
}
\email{botao@reality.design}

\author{Yilan Elan Tao}\authornotemark[1]
\orcid{0000-0003-1691-9727}
\affiliation{%
 \institution{Simon Fraser University}
 \city{Burnaby}
 \country{Canada}
}
\email{elan_tao@sfu.ca}

\author{Bernhard Riecke}
\orcid{0000-0001-7974-0850}
\affiliation{%
 \institution{Simon Fraser University}
 \city{Burnaby}
 \country{Canada}
}
\email{ber1@sfu.ca}

\author{Yue Li}
\orcid{0000-0003-3728-218X}
\affiliation{%
  \institution{Xi'an Jiaotong-Liverpool University}
  \city{Suzhou}
  \country{China}
  }
\email{yue.li@xjtlu.edu.cn}

\begin{abstract}
Mixed reality systems support shared anchors and co-located interaction, yet they lack a socially legible protocol for entering another person's mixed reality in public settings. We frame this as a protocol problem: co-located MR sharing requires a staged sequence---Discover, Consent, Confirm, Allow, Spatial Colocation, Sync Objects, Permission Management---each demanding user understanding and agreement. Using AirDrop and Apple Vision Pro SharePlay as a baseline, we show that MR encounter complexity far exceeds file transfer, yet must feel equally effortless. We present TouchPort, an embodied sharing protocol that collapses this multi-stage sequence into a single gesture: a handshake and pull that simultaneously signals intent, negotiates consent, and initiates a temporary shared encounter layer between otherwise separate mixed realities. Through three implied scenarios, we demonstrate the protocol's expressive range in the transition from isolated to spontaneously shared realities. We discuss how embodied gestures can address the consent problem in ubiquitous MR and examine the ethical tensions of encounter protocols for MR futures.
\end{abstract}




\begin{teaserfigure}
    \centering
        \includegraphics[width=0.9\linewidth]{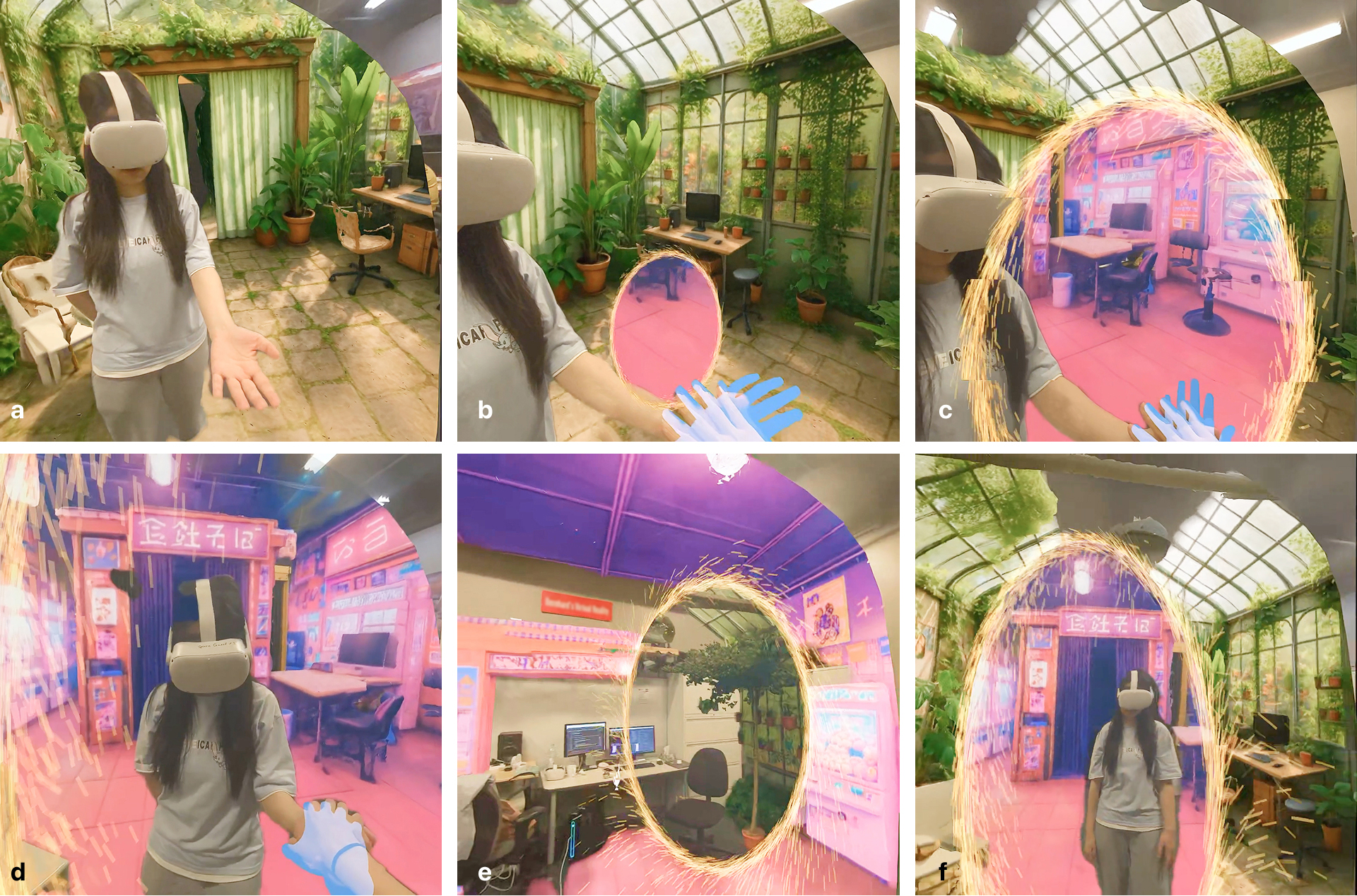}
    \caption{TouchPort: an embodied handshake-and-pull protocol for sharing mixed realities during spontaneous encounters. Each wearer inhabits a different mixed reality. (a) From the first-person perspective, the user sees their own green-themed room and a friend approaching. (b) Upon grasping the friend's hand (handshake), a portal emerges, offering a glimpse into the friend's mixed reality (pink-themed room). (c) The portal expands into a full magic circle, revealing the friend's MR perspective. (d) By gently pulling the user's hand, the friend guides them through the portal into her MR world. (e) The user's previous world remains spatially anchored on the far side of the portal, visible through the opening. (f) The user can return to their original MR environment by stepping back through the portal.}
    \label{fig:handshake}
    \Description{}
\end{teaserfigure}

\maketitle

\section{Introduction}

Mixed reality (MR) is arriving in public life~\cite{speicher2019mixed, grubert2017pervasive}. As headworn displays move from laboratories to sidewalks, caf\'{e}s, and transit~\cite{medeiros2023social}, people will increasingly occupy the same physical space while perceiving different digital layers---different digital objects, annotations, histories, and permissions overlaid on a shared material world~\cite{ens2019revisiting, hu2025autonomous}. Co-presence will no longer imply shared reality~\cite{medeiros2022shielding, lebeck2018}. Two people standing side by side may inhabit fundamentally different perceptual environments, each curated by their own devices, applications, and preferences~\cite{rajaram2025privacy}. The question of how one person enters another person's mixed reality---how a stranger, a colleague, or a friend might request, negotiate, and establish a temporary shared encounter---remains open~\cite{rajaram2023sharing}.

This gap reflects three compounding difficulties that current research has not addressed in concert. First, the problem of \textit{MR encounter establishment in the wild} is largely unstudied. The overwhelming majority of co-located MR collaboration research begins after connection is already established: participants are recruited, devices are paired, and a shared application is running~\cite{grasset2005, lukosch2015, dagan2022, mcgill2020, yu2022, hu2025}. The fragile, socially loaded moment \textit{before} sharing---where intent must be signaled, consent negotiated, scope defined, and trust calibrated---is absent from these accounts. Second, \textit{consent and privacy are deeply difficult in mixed reality}. Unlike screen-based sharing, MR involves embodied presence, spatial attention, and continuous environmental perception. Existing permission models---pop-up dialogs, toggle switches, one-time prompts---do not translate to a medium where sensing is always on and the boundary between private perception and shared space is invisible~\cite{roesner2014, denning2014, lebeck2017, lebeck2018, ohagan2022, abraham2024}. Bystanders cannot tell what is being sensed, shared, or recorded. Third, \textit{mixed reality encounters are naturally asymmetric}. Two people in the same physical space will never see exactly the same digital layer. Asymmetry is the default condition~\cite{yu2022, hoai2024, gugenheimer2017, numan2022}. Any sharing protocol must account for the fact that ``sharing'' does not mean ``seeing the same thing''---it means negotiating partial, bounded, and possibly unequal access to each other's augmented environments.

We argue that this is fundamentally a \textit{protocol problem}. What is missing is not a better rendering engine or a more intuitive menu, but a staged interaction sequence---Discover, Consent, Confirm, Allow, Colocation, Sync Objects, Permission Management---with each stage requiring user understanding and agreement. This protocol is an order of magnitude more complex than the nearest existing paradigm: cross-device file transfer. AirDrop moves a photo in three steps: discover nearby devices, send a request, and accept or reject it. An MR encounter requires negotiating temporary entry into a situated perceptual-computational layer, including spatial alignment, object persistence, permission scoping, ongoing management, and a graceful exit. Currently, Apple Vision Pro SharePlay provides the first nearby-sharing commercial solution of its kind. Yet moving through these stages remains complex and is not embodied intuitively.

We present TouchPort, an embodied sharing protocol that collapses the full encounter sequence into a single intuitive gesture: a handshake and pull that simultaneously signals intent, negotiates consent, confirms scope, and initiates a temporary shared encounter layer between otherwise separate mixed realities. The handshake maps to discovery and bilateral consent---it is inherently reciprocal, socially legible, and impossible to perform alone. The pull maps to threshold-crossing: a dramatized transition from isolated to shared reality that provides the system with a technical trigger for spatial alignment.

This paper makes four contributions: (1)~We frame MR encounter as a multi-stage protocol problem, using AirDrop and Apple Vision Pro SharePlay as baselines to expose the complexity gap. (2)~We present TouchPort, an embodied sharing protocol that merges discovery, consent, confirmation, and initiation into a single gesture interaction. (3)~We develop three implied scenarios---sharing past memories via Gaussian splatting, passing objects between worlds, and a public/private reality layer---that demonstrate the protocol's expressive range. (4)~We discuss design implications for embodied consent, spontaneous encounter, and the ethics of shared reality.

\section{Related Work}

\subsection{Co-located MR Encounters, Privacy, and Asymmetry}

Multi-user co-located mixed reality has been a research concern for nearly three decades~\cite{benford1993, billinghurst1998, szalavari1998, sereno2022}. \citet{szalavari1998} built Studierstube, one of the earliest collaborative AR environments. \citet{billinghurst1998} introduced Shared Space, framing AR as a medium for computer-supported cooperative work, and \citet{billinghurst2002} demonstrated that shared AR could serve as a face-to-face collaboration medium. Comprehensive surveys  \cite{lukosch2015,sereno2022,ens2019revisiting,schafer2023,radu2021} catalogue the field, mapping collaboration dimensions---symmetry, co-location, synchrony---and the evolution from CSCW groupware to modern MR. Recent work demonstrates playful co-located AR~\citep{dagan2022}, social MR in public settings~\citep{mcgill2020, nijholt2021}, coordination challenges in shared AR~\citep{chen2021}, and asymmetric co-located AR where HMD wearers and non-wearers share the same space~\citep{jansen2020}. \citet{hartmann2019} blend VR with physical surroundings, enabling awareness of co-located people while immersed. \citet{ritter2024} study trust negotiation in shared AR. \citet{hu2023mofa} has explored spontaneous co-located MR in public spaces: studied asymmetric MR multiplayer between handheld and head-mounted devices~\citep{hu2023mofa}, demonstrated that co-located MR street play with strangers is socially viable in diverse public settings~\citep{hu2025}, and developed instant spatial anchor sharing for ad-hoc session setup~\citep{hu2023instantcopresence}. This body of work establishes that spontaneous co-located MR is technically feasible and socially interesting even in uncontrolled public contexts. However, these studies operate within pre-structured game contexts where roles (e.g., wizard vs.\ muggle) are predefined and the shared experience is a single application. They do not address the general-purpose protocol problem: how two co-located people with no shared application context---a stranger, a colleague, or a friend---can initiate, negotiate, and establish a temporary shared encounter across otherwise separate mixed realities. Critically, even the most comprehensive survey~\citep{sereno2022} taxonomizes what happens \textit{during} a collaborative session but does not address how sessions begin or end. The fragile moment of establishing an encounter---discovering a nearby person, signaling intent, negotiating consent, and crossing the threshold from isolated to shared reality---remains unaddressed.

Research on AR privacy demonstrates that always-on sensing disrupts traditional permission metaphors. \citet{denning2014} found bystanders confused by AR glasses in public. \citet{ohagan2022} argue that everyday AR requires contextual, ongoing, and revocable mechanisms. \citet{lebeck2017, lebeck2018} propose output-level security for multi-user AR. \citet{roesner2014} call for new access-control models. \citet{abraham2024} show that capture LEDs are insufficient privacy indicators. \citet{corbett2023} implement bystander-data protections. \citet{regenbrecht2022} examine the ethical dimensions of pervasive AR, including bystander consent in public spaces. Nissenbaum's contextual integrity framework~\citep{nissenbaum2009} and Solove's taxonomy of privacy harms~\citep{solove2006} provide complementary theoretical bases for why MR permissions cannot follow simple notice-and-consent models. \citet{homayouni2023} propose consent-based design for social XR, and \citet{khan2025} explore making MR interfaces visible to bystanders. \citet{rajaram2023sharing} elicit user-preferred gestures and metaphors for sharing AR content, \citet{rajaram2023reframe} provide a storyboarding tool for analyzing security and privacy in multi-user AR scenarios, and \citet{rajaram2025privacy} propose a framework for dynamically balancing privacy needs as users enter and leave shared AR. Outside XR, \citet{zhou2024privacy} demonstrate interactive negotiation for privacy settings of co-located shared sensing devices---the closest analog to encounter negotiation, but in the IoT domain rather than MR. Together, this body of work identifies \textit{what} should be negotiated (object visibility, access scope, privacy boundaries) and \textit{how users think} about these concerns. However, no existing work proposes a concrete, end-to-end interaction protocol that unifies discovery, consent, session establishment, and graceful exit into a single legible act. TouchPort addresses this gap: it operationalizes the privacy and sharing concerns that prior work surfaces into a staged encounter protocol collapsed into an embodied gesture.

MR is inherently asymmetric: users occupy different positions along the reality-virtuality continuum~\citep{milgram1994}, each perceiving a distinct perceptual layer mediated by their own device and configuration. \citet{yu2022} demonstrate spatial conflicts in co-located AR. \citet{hoai2024} study public MR viewing in asymmetric groups. \citet{gugenheimer2017} explore cross-reality sharing with ShareVR. \citet{numan2022} investigate asymmetric collaborative MR. \citet{zaman2023} propose a taxonomy for mixed reality multi-user asymmetric collaboration. \citet{thoravikumaravel2020} bridge asymmetric VR communication. The concept of \textit{transitional interfaces}---systems that let users move fluidly along the reality-virtuality continuum---originates with the MagicBook~\citep{billinghurst2001magicbook} and was formalized by \citet{grasset2006transitional}. \citet{simeone2015substitutional} address physical-virtual pairing during such transitions. \citet{cools2022, cools2025} extend this lineage to cross-reality interaction techniques and design patterns. Recent surveys confirm the field remains sparse: \citet{auda2023crossreality} scope cross-reality systems broadly, while \citet{schroder2023collaborating} study how dyads collaborate when positioned at different virtuality levels---yet they do not address how such shared sessions are initiated or negotiated between co-located strangers. Asymmetry is not a bug but the fundamental condition of MR encounter. Yet no protocol addresses how asymmetric encounter should be initiated, scoped, and managed.

\subsection{Cross-Device Transfer Protocols and Embodied Interaction}

The closest paradigm for ``share something with a nearby person'' is cross-device transfer. \citet{rekimoto1997} introduced Pick-and-Drop. \citet{hinckley2003} demonstrated bumping phones to share data. \citet{lucero2011} studied embodied phone-to-phone sharing. \citet{nielsen2014} proposed JuxtaPinch for co-located pairing. \citet{hamilton2014} explored embodied transfer metaphors. \citet{voelker2020} demonstrate calibration-free cross-device collaboration through gaze and touch. \citet{marquardt2021} extend cross-device interaction to spatially-aware in-air device formations. \citet{emmert2024} present recent cross-device transfer advances. Consumer systems---AirDrop~\citep{apple2024airdrop}, Quick Share (formerly Nearby Share)~\citep{google2024nearbyshare}, and the now-defunct Bump---implement a lightweight protocol: discover, request, accept, transfer. These systems solve a simpler version of the encounter problem. But they transfer \textit{files}. MR encounter requires transferring \textit{world access}---spatial alignment, object persistence, permission scoping, and embodied co-presence that no file-transfer protocol addresses.

Proxemics research establishes that interpersonal distance, orientation, and gesture are socially meaningful. \citet{hall1966} defined foundational distance zones. \citet{ballendat2010} and \citet{greenberg2011} translated proxemics into HCI. \citet{gronbaek2020} extended proxemic interaction to public displays. \citet{marquardt2012} explored cross-device proxemic interaction. \citet{vogel2004} defined implicit-to-explicit interaction transitions. \citet{huang2022} applied embodied spatial interaction in AR. Most relevant to our work, \citet{lunding2023} use proxemics as a lens for designing handheld collaborative AR, where physical proximity triggers different collaboration modes---the closest existing work to using embodied spatial relationships as an initiation mechanism. However, their system is phone-based and assumes a shared application context rather than negotiating entry into another person's MR. Social acceptability research warns that public gestures carry stigma and cultural meaning~\citep{rico2010, montero2010, profita2013, koelle2020, alallah2018}. Any embodied initiation gesture for MR encounter must be socially legible, culturally navigable, and acceptable to participants and bystanders.

Neither cross-device transfer nor embodied interaction addresses the full complexity of MR encounter. While prior work has addressed individual facets---privacy negotiation~\citep{rajaram2023sharing, rajaram2025privacy}, security foundations~\citep{lebeck2018, deguzman2020}, co-located spatial alignment~\citep{mcgill2020}, asymmetric sharing~\citep{gugenheimer2017}, and proximity-triggered collaboration~\citep{lunding2023}---no work has proposed a unified interaction protocol spanning the full encounter lifecycle from discovery through consent, establishment, management, and graceful exit. AirDrop moves a photo; TouchPort must negotiate temporary entry into a situated perceptual-computational layer, and it must do so through a single legible gesture.

\section{Design Space: State Transitions in Existing MR Encounter Protocols}

To map the design space, we decompose three commercially deployed sharing protocols---Apple AirDrop, Apple Vision Pro SharePlay, and Meta Quest Colocation---into their constituent technical steps, examining what information each step requires and what state transition it produces. This decomposition reveals both the necessary stages of any encounter protocol and the specific gaps that motivate TouchPort.

\subsection{AirDrop: File Transfer as Minimal Encounter}

AirDrop~\citep{apple2024airdrop} is the simplest widely-deployed protocol for sharing with a nearby person. Its state machine has seven transitions (no Spatial Alignment is needed):

\begin{enumerate}
\item \textbf{Prerequisite} (\textit{isolated} $\to$ \textit{eligible}). Both sender and receiver must have AirDrop enabled (set to ``Contacts Only'' or ``Everyone''), with Wi-Fi and Bluetooth turned on. Devices must be within approximately 9~meters. \textit{Information required:} system settings, radio hardware. \textit{Cost:} enabling ``Everyone'' mode exposes the device to unsolicited transfers from strangers.
\item \textbf{Discovery} (\textit{eligible} $\to$ \textit{aware}). The sender's device broadcasts truncated SHA-256 hashes of the sender's Apple~ID and phone number via BLE. Nearby devices compare these hashes against their contact lists to resolve identity. \textit{Information required:} Apple~ID, phone number, BLE radio. \textit{Privacy cost:} identity hashes leak to \textit{all} nearby devices; Heinrich et al.~\cite{heinrich2021} showed these are crackable via rainbow table in under one second.
\item \textbf{Intent} (\textit{aware} $\to$ \textit{requested}). The sender selects content and taps a recipient from the discovered device list. A peer-to-peer Wi-Fi connection is established. \textit{Information required:} content selection, recipient choice.
\item \textbf{Preview} (\textit{requested} $\to$ \textit{previewed}). The receiver sees a file preview---a thumbnail of the incoming content. AirDrop is the only baseline that offers any preview, though it is bundled into the consent dialog. \textit{Information required:} content thumbnail.
\item \textbf{Consent} (\textit{previewed} $\to$ \textit{accepted}). The receiver taps accept or reject---binary, no scope negotiation. \textit{Privacy cost:} cyberflashing---unsolicited image transmission---is now a criminal offense under the UK's Online Safety Act 2023.
\item \textbf{Sync} (\textit{accepted} $\to$ \textit{sharing}). The file copies unidirectionally. No spatial alignment is needed.
\item \textbf{Termination} (\textit{sharing} $\to$ \textit{isolated}). The connection terminates. No shared state persists.
\end{enumerate}

AirDrop succeeds because what is shared---a file---is simple, bounded, and unidirectional. The protocol requires no spatial alignment, no ongoing synchronization, and no permission management beyond the initial accept.

\subsection{Apple Vision Pro SharePlay: Same-Room MR Sharing}

SharePlay is the most advanced commercial system for co-located MR sharing. With visionOS~26, announced at WWDC25~\citep{apple2025shareplay}, Apple extended SharePlay to support same-room encounters. Its state machine has eight transitions:

\begin{enumerate}
\item \textbf{Prerequisite} (\textit{isolated} $\to$ \textit{eligible}). Both users must have an Apple~ID, FaceTime account, and Vision~Pro running visionOS~26 with the \textit{same app} installed. \textit{Information required:} platform identity, app parity. \textit{Cost:} ecosystem lock-in excludes $\sim$85\% of global mobile users.
\item \textbf{Discovery} (\textit{eligible} $\to$ \textit{aware}). Devices detect each other via BLE and peer-to-peer Wi-Fi. \textit{Information required:} BLE/Wi-Fi radio, proximity.
\item \textbf{Intent} (\textit{aware} $\to$ \textit{requested}). One user sends a SharePlay invitation, signaling a desire to share. \textit{Information required:} sender identity, app context. The intent is unilateral---only one party initiates.
\item \textbf{Preview}. N/A---there is no preview stage. The recipient cannot inspect what will be shared before deciding.
\item \textbf{Consent} (\textit{requested} $\to$ \textit{accepted}). The recipient receives a binary ``Join?'' system dialog. Accept or reject. \textit{Information required:} none beyond identity. \textit{Limitation:} consent is all-or-nothing with no opportunity to negotiate scope.
\item \textbf{Spatial Alignment} (\textit{accepted} $\to$ \textit{co-located}). ARKit automatically establishes a shared coordinate system. \textit{Information required:} visual-inertial odometry, environmental feature points.
\item \textbf{Sync} (\textit{co-located} $\to$ \textit{sharing}). GroupSessionMessenger synchronizes app-level state between devices. Both enter a shared immersive space. \textit{Information required:} app state, object transforms.
\item \textbf{Termination} (\textit{sharing} $\to$ \textit{isolated}). Exit through a system menu. No graceful ritual, no residue management. Binary leave.
\end{enumerate}

SharePlay adds spatial alignment and synchronization to the state machine but offers no mechanism for stranger interaction, no scope negotiation, and all-or-nothing consent.

\subsection{Meta Quest Colocation: Cloud-Mediated Spatial Sharing}

Meta Quest Colocation~\citep{meta2024colocation} provides Meta's approach via the Colocation Discovery API (SDK~v71+). Its state machine has eight transitions:

\begin{enumerate}
\item \textbf{Prerequisite} (\textit{isolated} $\to$ \textit{eligible}). All users must own a Meta Quest headset with accounts and ``Share point cloud data'' enabled. \textit{Information required:} Meta account, privacy consent for cloud processing.
\item \textbf{Discovery} (\textit{eligible} $\to$ \textit{aware}). One device advertises a session with a unique group UUID via Bluetooth ($\sim$10m range). Others discover and list available sessions. \textit{Information required:} Bluetooth radio, group UUID.
\item \textbf{Intent} (\textit{aware} $\to$ \textit{requested}). One user creates a colocation group and broadcasts an invitation. \textit{Information required:} group UUID, session metadata. The intent is unilateral---only the host initiates.
\item \textbf{Preview}. N/A---there is no preview stage. Joiners cannot inspect what will be shared.
\item \textbf{Consent} (\textit{requested} $\to$ \textit{accepted}). Others join by accepting the invitation. Binary consent---no scope negotiation. \textit{Information required:} group UUID.
\item \textbf{Spatial Alignment} (\textit{accepted} $\to$ \textit{co-located}). Shared Spatial Anchors: point cloud data from each headset's camera-based tracking is routed through Meta's cloud servers, where algorithms find matching environmental features to align virtual coordinate systems. \textit{Information required:} environment point clouds. \textit{Privacy cost:} spatial data leaves the device.
\item \textbf{Sync} (\textit{co-located} $\to$ \textit{sharing}). Single-app state synchronization within a single application context. \textit{Information required:} app state.
\item \textbf{Termination} (\textit{sharing} $\to$ \textit{isolated}). Leave session. Binary exit.
\end{enumerate}

Meta Quest Colocation improves on SharePlay by offering group-based UUID sharing that eliminates per-user entitlement checks, but introduces a cloud dependency for spatial alignment. It shares the same fundamental limitations: no stranger encounters, no embodied initiation, binary consent, single-app confinement, and no concept of asymmetric or public layers.

\subsection{The Design Space Gap}

Decomposing these three protocols reveals eight necessary stages in any encounter protocol (Table~\ref{tab:comparison}). Navigating all eight stages is complex and time-consuming: even SharePlay's streamlined flow requires multiple menu interactions, system dialogs, and waiting periods before two people in the same room can share an experience. Meta Quest Colocation adds cloud round-trips for spatial alignment. In practice, these friction costs mean that co-located MR sharing remains a deliberate, pre-arranged activity rather than something that happens spontaneously during everyday encounters.

The problem intensifies when we consider that MR encounters are inherently \textit{asymmetric} and \textit{transient}. In a future of pervasive MR, a person walking through a city may want to discover, enter, and exit multiple different shared realities in quick succession---glancing into a street performer's augmented stage, briefly sharing a memory with a friend at a caf\'{e}, picking up a digital object left at a landmark. Each of these encounters involves a different counterparty, different content, different permissions, and different duration. The eight-stage protocol must be traversed each time. If each traversal requires the overhead of current systems---account verification, app matching, menu navigation, binary consent dialogs, cloud-mediated alignment---spontaneous encounter is effectively impossible. The protocol is necessary, but it must become invisible.

\begin{table*}[t]
\caption{State transitions across AirDrop, Vision Pro SharePlay, Meta Quest Colocation, and TouchPort. Each row is a protocol stage; each cell shows the state transition (\textit{from} $\to$ \textit{to}) and the information required.}
\label{tab:comparison}
\small
\begin{tabularx}{\textwidth}{@{}l l X X X X@{}}
\toprule
\textbf{Stage} & \textbf{Transition} & \textbf{AirDrop} & \textbf{Vision Pro SharePlay} & \textbf{Meta Quest Colocation} & \textbf{TouchPort} \\
\midrule
Prerequisite & isolated $\to$ eligible & AirDrop, Wi-Fi, Bluetooth enabled & Apple~ID, FaceTime, same app, visionOS~26 & Meta account, ``Share point cloud'' enabled & MR headset with hand tracking \\
Discover & eligible $\to$ aware & BLE broadcast of identity hashes (leaks PII) & BLE + peer-to-peer Wi-Fi & Bluetooth group UUID ($\sim$10m) & BLE presence beacons; no identity exchanged \\
Intent & aware $\to$ requested & Sender selects recipient & One user sends SharePlay invitation & Host creates colocation group & Extend hand (unilateral signal) \\
Preview & requested $\to$ previewed & File preview (bundled with consent) & N/A (all-or-nothing) & N/A (all-or-nothing) & Scope preview: objects, layers, duration, rights \\
Consent & previewed $\to$ accepted & Receiver accept/reject (binary) & ``Join?'' prompt (binary) & Accept invite (binary) & Reciprocal handshake (bilateral) \\
Spatial Align & accepted $\to$ co-located & N/A & ARKit shared coordinates & Cloud-mediated spatial anchors & On-device alignment; mismatch visible \\
Sync & co-located $\to$ sharing & Unidirectional file copy & GroupSessionMessenger & Single-app state sync & Move, copy, lend, mirror, or co-own \\
Terminate & sharing $\to$ isolated & Transfer complete & Leave via menu (binary) & Leave session (binary) & Graceful release with residue management \\
\bottomrule
\end{tabularx}
\end{table*}

Moreover, MR encounter requires transporting three interdependent kinds of information that file-transfer protocols never face: (1)~\textit{colocation data}---shared spatial anchors or point clouds to align independent world models; (2)~\textit{privacy and consent state}---granular, bilateral, and ongoing negotiation over what each party reveals and withholds; and (3)~\textit{preview representations}---lightweight descriptions of the proposed shared experience so users can make informed consent decisions. These substrates are coupled: colocation data carries privacy implications (point clouds reveal room geometry), consent constrains what colocation data can be exchanged, and preview requires partial colocation to show spatially grounded content. The core design challenge is: can we collapse this multi-stage, multi-substrate negotiation into a single embodied gesture---making the necessary protocol invisible while preserving its full expressive power?

\section{Our Design: TouchPort}

\subsection{The Core Idea}

TouchPort collapses the multi-stage encounter sequence into a single gesture: a handshake and pull. The \textbf{handshake} accomplishes multiple protocol stages simultaneously. Extending a hand signals encounter intent (Discover). Reciprocation constitutes bilateral consent (Consent)---the handshake cannot be performed unilaterally. At contact, a brief scope preview appears to both parties (Confirm). The \textbf{pull}---a gentle drawing motion---triggers encounter initiation (Allow), provides a directional metaphor for threshold-crossing, and gives the system a technical trigger for spatial alignment (Colocation). The multi-stage protocol happens \textit{within} the gesture, not before it.

This design follows a lineage of embodied transfer metaphors: Pick-and-Drop~\cite{rekimoto1997}, Bump~\cite{hinckley2003}, JuxtaPinch~\cite{nielsen2014}, and ShakeCast (Weissker et al., 2017), which demonstrated handshake detection for automated contact exchange. TouchPort extends this lineage from transferring \textit{content} to negotiating temporary \textit{world access}.

Critically, the collapse does not eliminate protocol stages---it \textit{compresses} them into a denser signal. Table~\ref{tab:collapse} maps each stage to its embodied realization.

\begin{table}[h]
\caption{Protocol stage collapse: sequential vs.\ embodied.}
\label{tab:collapse}
\small
\begin{tabular}{@{}p{1.6cm}p{2.8cm}p{2.4cm}@{}}
\toprule
\textbf{Stage} & \textbf{Traditional} & \textbf{TouchPort} \\
\midrule
Discover & BLE scan $\rightarrow$ device list $\rightarrow$ select & Ambient cue (auto) \\
Consent & Screen notification $\rightarrow$ tap Accept & Hand extension $\rightarrow$ clasp \\
Confirm & Read dialog $\rightarrow$ confirm & Scope preview during grip \\
Allow & Tap ``Start'' / ``Join'' & Pull gesture \\
Colocation & Automatic (hidden) & Handshake midpoint = anchor \\
\midrule
\multicolumn{2}{l}{\textbf{Traditional:}} & 5 screens, $\sim$15--30s \\
\multicolumn{2}{l}{\textbf{TouchPort:}} & 2 gestures, $\sim$3--5s \\
\bottomrule
\end{tabular}
\end{table}

\subsection{Design Rationale}

\textbf{Why one gesture instead of many steps.} Protocol complexity should be absorbed by the system, not imposed on the user. AirDrop succeeded by hiding its protocol behind one tap. Embodied interaction carries more social meaning per unit time than screen-based dialogs~\cite{ballendat2010, greenberg2011}.

\textbf{Why embodied instead of screen-based.} MR encounters are bodily. A gesture is visible to participants and bystanders simultaneously---serving as both consent mechanism and social indicator~\cite{rico2010, koelle2020}. Screen-based consent breaks immersion and communicates nothing to nonparticipants.

\textbf{Why the handshake.} It is among the most universally recognized gestures of mutual agreement. It carries built-in reciprocity, dense cultural meaning (greeting, partnership, trust)~\cite{hall1966}, and creates a physical anchor for the digital transition. The handshake also provides security properties that digital protocols cannot match: (1)~\textit{physical co-presence proof}---unlike BLE broadcast (range $\sim$10m, easily relayed), the handshake requires bodies within arm's reach ($\sim$0.5m), verified by independent hand tracking on both devices; (2)~\textit{mutual attention proof}---both parties must be facing each other, eliminating AirDrop's ``accept while distracted'' attack surface; (3)~\textit{bilateral action proof}---accidentally extending your hand, clasping someone else's, and sustaining grip requires 3+ deliberate motor actions, unlike a single accidental notification tap; (4)~\textit{non-replayability}---a handshake is a live physical event that cannot be captured and replayed like a digital authentication token.

\textbf{Limitations.} The handshake is culturally specific, not universally comfortable~\cite{rico2010, montero2010}, requires two functional hands, and carries post-pandemic hygiene sensitivity. We present it as one exemplar in a family of possible initiation gestures; alternatives are discussed in the appendix.

\begin{figure*}
    \centering
        \includegraphics[width=1\linewidth]{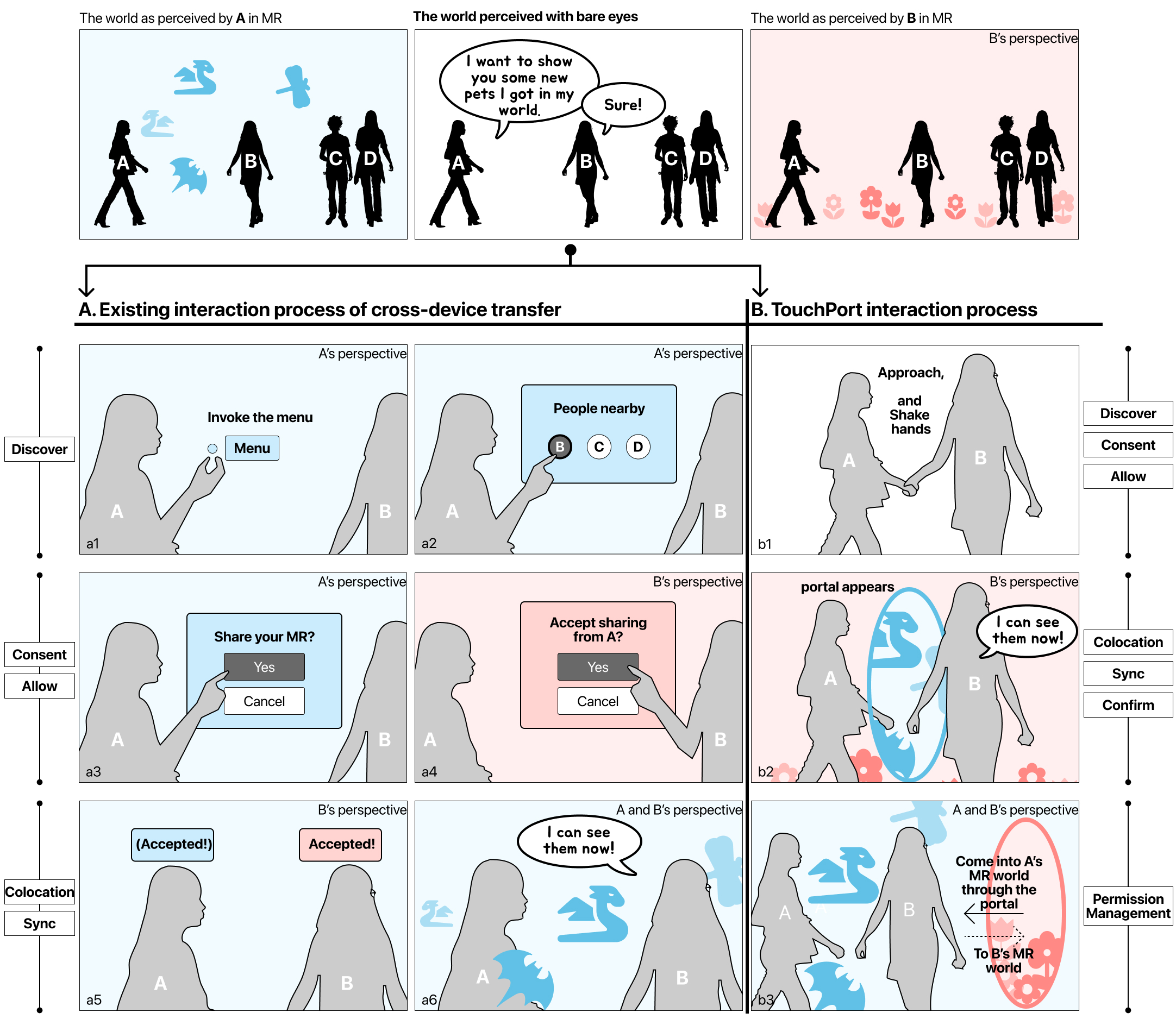}
    \caption{Comparing the interaction process of mixed reality sharing between (A) the existing cross-device transfer protocols and (B) TouchPort. }
    \label{fig:process}
    \Description{}
\end{figure*}

\subsection{Interaction Flow}

TouchPort collapses the eight-stage protocol (Table~\ref{tab:comparison}) into five embodied phases (Figure~\ref{fig:handshake}). Each phase merges one or more protocol stages into a continuous physical action:

\begin{enumerate}
\item \textbf{Approach.} Walking toward another person already accomplishes two protocol stages at once. Devices detect each other via BLE beacons, and the act of approaching signals social intent---just as in everyday life, moving toward someone communicates interest before any words are exchanged.\\
$\to$ \textit{Prerequisite} (isolated $\to$ eligible) + \textit{Discovery} (eligible $\to$ aware) + \textit{Intent} (aware $\to$ requested).

\item \textbf{Handshake.} Upon grasping the other person's hand, a preview portal emerges at the point of contact---a small window into the other person's mixed reality (Figure~\ref{fig:handshake}b). Both parties can see a glimpse of what the other's world contains: objects, environment, and scope. Either can release to abort.\\
$\to$ \textit{Preview} (requested $\to$ previewed) + \textit{Consent} (previewed $\to$ accepted). The handshake is inherently bilateral---it cannot be performed unilaterally.

\item \textbf{Pull.} A gentle pull expands the preview portal into a full passage between realities (Figure~\ref{fig:handshake}c--d). Spatial alignment occurs between both devices' world models. The shared encounter layer instantiates with agreed-scope objects.\\
$\to$ \textit{Spatial Alignment} (accepted $\to$ co-located) + \textit{Sync} (co-located $\to$ sharing).

\item \textbf{Explore.} Both participants can now move through the shared space. The user's previous world remains spatially anchored on the far side of the portal, visible through the opening (Figure~\ref{fig:handshake}e). Permission management is ongoing---either participant can escalate, downgrade, or selectively revoke.

\item \textbf{Release.} Either participant steps back through the portal or releases the shared layer. Shared objects fade per agreed residue terms. Both return to their private realities (Figure~\ref{fig:handshake}f).\\
$\to$ \textit{Termination} (sharing $\to$ isolated).
\end{enumerate}

The key insight is that the handshake-and-pull compresses all eight protocol stages into a single continuous physical sequence. Approach merges Prerequisite, Discovery, and Intent. The handshake merges Preview and Consent: grasping opens the preview portal, and sustaining the grip constitutes bilateral consent. The pull merges Spatial Alignment and Sync: the physical motion provides both the trigger and the directional metaphor for crossing into another person's reality.

\subsection{Permission Management During Encounter}

After initiation, the protocol continues. We identify eight permission dimensions:

\begin{table}[h]
\small
\begin{tabular}{@{}ll@{}}
\toprule
\textbf{Dimension} & \textbf{What It Governs} \\
\midrule
Scope & What can be perceived \\
Action & What can be done (view, manipulate) \\
Duration & How long access persists \\
Audience & Who else can access \\
Mutability & Whether shared objects can be changed \\
Portability & Whether objects can leave the encounter \\
Revocability & How quickly access withdraws \\
Residue & What remains after encounter ends \\
\bottomrule
\end{tabular}
\end{table}

These dimensions draw on AR security work~\cite{lebeck2018, ohagan2022, abraham2024} but extend it from app-level access control to interpersonal encounter governance. Permissions are negotiated in layers: defaults at initiation, adjustable during encounter through lightweight gestural controls. Full specification in Appendix~B.

\section{Implementation}

We implemented TouchPort as a proof-of-concept prototype. We describe each component and its implementation status.

\textbf{Proximity Detection.} The Discover stage uses BLE advertising with rotating, privacy-preserving encounter tokens. Two Meta Quest~3 headsets within approximately 10~meters detect each other via Bluetooth and trigger subtle ambient cues. Fully implemented using the Meta Quest Bluetooth sharing protocol.

\textbf{Hand Tracking and Handshake Detection.} We detect handshakes using the Meta Hand Tracking SDK on Meta Quest~3. The pipeline identifies: (1)~two hands from different devices within 15~cm, (2)~open-palm-forward orientation, (3)~clasp contact, (4)~sustained grip for $\geq$0.8~seconds. Fully implemented; detection degrades in bright outdoor lighting.

\textbf{Spatial Anchor Alignment.} At pull completion, devices exchange anchor data using Meta Shared Spatial Anchors via the Colocation Discovery API. The shared coordinate origin is established at the handshake midpoint. Alignment completes in under 500~ms for same-room encounters. Fully implemented.

\textbf{Object Synchronization.} A peer-to-peer data channel carries object state at 30~Hz. Three transfer semantics are supported: \textit{move} (sender loses object), \textit{copy} (both retain), and \textit{lend} (reverts on termination). Move semantics use two-phase commit for atomicity. Fully implemented for geometric primitives and glTF models.

\textbf{Permission Token System.} Scoped, time-limited, revocable JSON tokens encode permissions. Revocation is immediate via gesture or distance threshold. Token specification fully implemented. User-facing permission negotiation UI is currently Wizard-of-Oz.

\textbf{Gaussian Splatting Integration.} 3D Gaussian splatting~\cite{kerbl2023} enables memory sharing (\S6.1). Splat assets are spatially anchored and streamed peer-to-peer. Rendering achieves 45--60~fps for $\leq$500K Gaussians on Meta Quest~3.

\textbf{Bystander Indicators.} Designed (pulsing LED ring, audible chime) but not yet implemented.

\section{Application Scenarios}

We outline three future applications that demonstrate how the TouchPort protocol extends beyond a single interaction technique to support qualitatively different forms of MR sharing.

\subsection{Sharing Memories}

TouchPort enables sharing volumetric captures---such as 3D Gaussian splats~\cite{kerbl2023}---as spatially situated, permission-managed encounters. One user can share a captured place (e.g., a demolished family garden) with a co-located person via the handshake-and-pull protocol. The preview portal shows a thumbnail of the scene with duration and permission constraints (view only, no recording). Upon consent, the splat renders as a navigable spatial overlay anchored between both users. Permission tokens enforce scope: the receiver cannot capture, record, or retain the memory. The encounter terminates automatically when the agreed duration expires.

This application extends the cross-device transfer lineage from files to \textit{situated spatial memories}~\cite{sellen2010, pejsa2016, beck2020}. Unlike sending a photograph, sharing a volumetric capture is inherently asymmetric---the sharer controls scope, duration, and rights. The protocol supports this asymmetry by design through scoped permission tokens.

\subsection{Passing Objects Between Worlds}

TouchPort supports transferring virtual objects between users' independent MR environments. The protocol handles three transfer semantics: \textit{move} (atomic removal from sender, instantiation in receiver, via two-phase commit)~\cite{rekimoto1997, hamilton2014, emmert2024}, \textit{copy} (both retain a version, with permission tokens encoding modification and redistribution rights), and \textit{lend} (object automatically returns to sender when the encounter ends). Object transfer occurs at the spatial midpoint established during the pull phase. This extends the cross-device transfer lineage~\cite{cools2022} from file-level portability to world-level interoperability (Figure~\ref{fig:publiclayer}).

\begin{figure*}
    \centering
        \includegraphics[width=1\linewidth]{images/publiclayer_ownership.jpg}
    \caption{Private and public MR layers and object transfer between different worlds with TouchPort. Users may inhabit their private MR layer (top) or the public MR layer (bottom), with objects transferable directly between private worlds via a double-sided portal, or indirectly through the public layer---where deposited objects become unowned.}
    \label{fig:publiclayer}
    \Description{}
\end{figure*}

\subsection{Public and Private Reality Layers}

The protocol extends beyond dyadic encounters to support a \textit{public layer}: a shared, location-anchored MR commons where users can deposit and discover virtual objects (Figure~\ref{fig:publiclayer}). Objects placed in the public layer carry metadata---anonymous or attributed authorship, expiration time, pickup rules (single or unlimited), and embedded content visible only upon pickup. The protocol stages map differently here: the ``counterparty'' is the commons itself, consent is implicit in approaching, and permissions are encoded at placement time rather than negotiated in real time.

The public layer raises questions of \textit{curation} (what prevents virtual littering?)~\cite{colley2017}, \textit{anonymity vs.\ accountability} (anonymous placement enables gifting but also harassment), and \textit{persistence} (should MR commons accumulate history like physical graffiti?)~\cite{kljun2016, ryskeldiev2018, benford1998}. This extends encounter protocols from person-to-person to person-to-place relationships.

Together, these three applications demonstrate that the same core protocol primitives---discovery, consent, spatial alignment, scoped permission tokens, and graceful exit---support intimate memory exchange, interpersonal object transfer, and public spatial commons.

\section{Discussion}

\subsection{Embodied Gesture as Consent Mechanism}

The central design claim of TouchPort is that an embodied gesture can simultaneously serve as social signal, technical trigger, consent mechanism, and bystander indicator---collapsing what would otherwise be a multi-stage permission workflow into a single legible act.

Traditional consent mechanisms in computing operate through what we call \textit{declarative consent}: the user reads a description and clicks to affirm understanding. These mechanisms have well-documented failures even in screen-based contexts~\cite{nissenbaum2009}; in embodied MR, where interactions are spatial, continuous, and socially situated, they are fundamentally inadequate. A pop-up dialog interrupts immersion, communicates nothing to bystanders, and reduces nuanced social negotiation to a binary click. TouchPort proposes an alternative: \textit{performative consent}, where the act of consenting is itself the social and technical initiation of sharing. The handshake is not a metaphor for consent---it \textit{is} consent, enacted through the body.

We position TouchPort within the speculative design tradition~\cite{dunne2013, auger2013, elsden2017}, where designed artifacts serve not as finished products but as provocations that surface questions about possible futures. Following Elsden et al.'s speculative enactments~\cite{elsden2017} and Buchenau and Suri's experience prototyping~\cite{buchenau2000}, the scenarios in \S6 are structured provocations: they ask what it would mean to share a memory, hand over an object, or leave something in a public layer. By grounding these provocations in a specific gesture and protocol architecture, we move speculative design toward what Blythe calls ``research through design fiction''~\cite{blythe2014}---speculation precise enough to be critiqued, extended, and tested.

The deeper argument is that encounter protocols are infrastructure. Just as network protocols shaped the conditions of possibility for digital communication, MR encounter protocols will shape the social conditions of future mixed reality. Who can initiate? What can be shared? How is consent withdrawn? These are not UX details---they are political questions embedded in technical architecture. If MR is to become a medium for public life, the design of its encounter protocols deserves the same attention given to its rendering pipelines.

\subsection{From Isolated to Spontaneously Shared}

The default state of current mixed reality is isolation. Each user inhabits a private perceptual layer: their own spatial anchors, virtual objects, annotations, and overlays. Even when two MR users occupy the same physical space, they share geography but not reality. This is not a temporary limitation of early hardware---it is the architectural default. Without an encounter protocol, MR is solitary by design.

This matters because spontaneous encounter is the foundational mode of human social life. Goffman's analysis of behavior in public places~\cite{goffman1963} describes how strangers continuously negotiate co-presence through glances, body orientation, micro-gestures, and what he calls ``civil inattention''---the subtle acknowledgment that another person exists without demanding engagement. These encounters are unplanned, uninvited, and often fleeting, but they constitute the social fabric of public space. Simmel's foundational sociology~\cite{benford1993} similarly emphasizes that the encounter---the momentary coming-together of individuals who may never meet again---is the elemental unit of social life.

Current MR systems support only pre-arranged sharing: both users must open the same application, join the same session, and often be on the same platform. Apple Vision Pro's SharePlay requires a FaceTime call or mutual app context. This is the equivalent of requiring a formal appointment for every human interaction---it eliminates the caf\'{e} encounter, the park bench conversation, the chance meeting on a train. If MR cannot support unplanned sharing, it will remain a solitary medium even in the most social of physical spaces, producing what we call \textit{co-located loneliness}: the experience of being physically together but perceptually alone.

TouchPort is designed to restore the possibility of spontaneous encounter. The handshake-and-pull requires no prior arrangement, no shared app, and no platform coordination. Its minimal prerequisites---physical proximity and mutual willingness expressed through gesture---mirror the minimal prerequisites of face-to-face encounter in unmediated public space.

\subsection{Ethical Tensions}

We do not claim that TouchPort resolves the ethical challenges of MR encounter; rather, it surfaces them in concrete form. Several tensions deserve honest examination.

\textbf{Power asymmetry in initiation.} Any encounter protocol creates an asymmetry between initiator and responder. The person who extends their hand first sets the terms: they choose when, where, and toward whom the gesture is directed. While the handshake is reciprocal in form, the social dynamics of refusal are not symmetric. Declining a handshake carries social cost~\cite{ohagan2022, nissenbaum2009}, and in contexts where power differentials exist, the ``freedom to refuse'' may be nominal rather than real. Protocol design cannot eliminate social pressure, but TouchPort's requirement for mutual gesture completion means that passive non-response is itself a valid and low-cost refusal.

\textbf{Surveillance and tracking risk.} The pre-network discovery phase---where devices detect nearby MR-capable people---creates a new tracking surface. Even before any encounter is initiated, the system must sense who is nearby~\cite{denning2014, corbett2023}. Discovery must be designed with privacy-preserving mechanisms (ephemeral identifiers, local-only processing, no persistent logs), but the tension between discoverability and privacy is fundamental.

\textbf{Consent quality.} Is a gesture sufficient for informed consent? A handshake communicates willingness but not necessarily understanding. The scope preview is necessarily brief; users may not fully comprehend what they are agreeing to share~\cite{roesner2014, lebeck2018}. Embodied consent introduces temporal pressure that screen-based consent does not: the gesture is a real-time social act that resists the ``pause and read'' affordance of a dialog box.

\textbf{Cultural and accessibility exclusion.} The handshake is not culturally neutral. It carries specific meanings in Western contexts that do not translate universally; in some cultures, physical contact between strangers is inappropriate~\cite{rico2010, montero2010, profita2013}. The gesture also requires two functional hands. We present it as one exemplar in a family of possible initiation gestures, but the choice of gesture is itself a political act that privileges certain bodies and cultural norms.

\textbf{Bystander ethics.} Even with visible indicators, bystanders may not understand what is happening. Two people performing a handshake may appear engaged in an ordinary greeting; the fact that they are negotiating perceptual access is invisible to non-participants~\cite{denning2014, abraham2024}. What do non-users of a pervasive technology have a right to know about its operation in shared space?

\subsection{Limitations and Future Work}

This work has several important limitations. First, the handshake-and-pull is culturally specific and not universally accessible. Future work should explore culturally adaptive initiation gestures and investigate how different communities would design their own encounter rituals.

Second, TouchPort is presented as a design exploration and speculative protocol, not a deployed system. While individual technical components are grounded in existing capabilities, the full end-to-end protocol has not been subjected to user evaluation in realistic field conditions. Following Wong and Khovanskaya~\cite{wong2018}, we position this as a structured provocation, while acknowledging that empirical validation---through Wizard-of-Oz studies, field deployments, and longitudinal observation---is necessary to understand how people actually experience encounter protocols.

Third, the current design addresses only dyadic encounter. Group encounters, crowd-scale sharing, and multi-party permission management introduce combinatorial complexity. How does consent work when five people want to join? How are permissions negotiated where trust levels vary?

Fourth, the permission management framework identifies relevant dimensions but does not fully specify their implementation or evaluate their usability. How users understand and manipulate these dimensions during an active encounter---while simultaneously engaged in social and spatial demands---is a rich area for future investigation.

Finally, the public layer scenario opens governance questions that extend beyond interaction design: who moderates a public MR commons, how is virtual litter managed, and what are the property rights over persistent public objects? These connect to broader debates about digital commons and platform governance that this paper can only begin to frame.

\section{Conclusion}

Mixed reality is becoming a medium for everyday life, but it remains fundamentally private: each person perceives their own digital layer, and there is no shared ritual for crossing the boundary between one person's reality and another's. This paper frames co-located MR sharing as a protocol problem---a staged sequence of discovery, consent, confirmation, spatial alignment, and permission management that current systems leave unaddressed. We present TouchPort, an embodied sharing protocol that collapses this complexity into a single gesture: a handshake and pull that simultaneously signals intent, negotiates consent, and initiates a temporary shared encounter layer. Through three implied scenarios, we demonstrate the protocol's expressive range and surface the design questions that MR encounter demands. The design of encounter protocols is not a peripheral UX concern---it is infrastructure that will shape the social conditions of future mixed reality as profoundly as network protocols shaped the web. If mixed reality is to move from solitary tool to shared medium, it will need not only better rendering, but better rituals---rituals that make it possible, legible, and revocable to say: \textit{allow me into your dream}.

\bibliographystyle{ACM-Reference-Format}
\bibliography{references}

\end{document}